\documentclass[a4paper,twoside]{article}
\newcommand{\affil}[1]{$^{\rm #1}$}
\usepackage[authoryear]{natbib}
\usepackage{txfonts}
\usepackage{lmodern}
\bibpunct{(}{)}{;}{a}{}{,}
\usepackage{graphicx}
\usepackage{placeins}
\date{} %Please leave the date blank
\title{\large\bf\flushleft Simulating the Role of Stellar Rotation in the Spectroscopic Effects of Differential Limb Magnification}
\author{\parbox{\textwidth}{\flushleft
\vspace{-0.5cm}
{\it Adam R.~H. Stevens\affil{A,B,C}, Michael D. Albrow\affil{B}}\\
\vspace{0.4cm}
{\small \affil{A}\,Centre for Astrophysics \& Supercomputing, Swinburne University of Technology, Hawthorn, VIC 3122, Australia}\\
{\small \affil{B}\,Department of Physics and Astronomy, University of Canterbury, Private Bag 4800, Christchurch, New Zealand}\\
{\small \affil{C}\,Email: astevens@swin.edu.au}}}
\begin{document}
\twocolumn[
\begin{changemargin}{.8cm}{.5cm}
\begin{minipage}{.9\textwidth}
\vspace{-1cm}
\maketitle
%
%
%%%%%%%%%%%%%     ABSTRACT    %%%%%%%%%%%%%
%Abstract of no more than 200 words here.
\small{\bf Abstract: Finite-source effects of gravitationally microlensed stars have been well discussed in the literature, but the role that stellar rotation plays has been neglected.  A differential magnification map applied to a differentially Doppler-shifted surface alters the profiles of absorption lines, compromising their ordinarily symmetric nature.  Herein, we assess the degree to which this finite-source effect of differential limb magnification (DLM), in combination with stellar rotation, alters spectroscopically derived stellar properties.  To achieve this, we simulated a grid of high-magnification microlensing events using synthetic spectra.  Our analysis shows that rotation of the source generates differences in the measured equivalent widths of absorption lines supplementary to DLM alone, but only of the order of a few percent.  Using the wings of H$\alpha$ from the same simulated data, we confirmed the result of \citet{johnson10} that DLM alters measurements of effective temperature by $\lesssim$ 100 K for dwarf stars, while showing rotation to bear no additional effect.}

%%%%%%%%%%%%%     KEYWORDS    %%%%%%%%%%%%%
\medskip{\bf Keywords:} gravitational lensing: micro - line: profiles - stars: rotation
% Please write all keywords in lower case. PASA uses the
% standard list of subject headings adopted by The Astrophysical Journal
% and available from http://www.journals.uchicago.edu/ApJ/keywords_text.html.
% Keywords are separated by em-dashes, i.e. ---

%%%%%%%%DO NOT EDIT%%%%%%%%%%%%
\medskip
\medskip
\end{minipage}
\end{changemargin}
]
\small
%%%%%%%%EDIT FROM HERE%%%%%%%%%%%%

\section{Introduction}
\label{sec:intro}
%Please see the PASA Style Guide for help with correct layout for your manuscript.
%Examples of tables and figures are given below.

Gravitational microlensing occurs when an object of sufficient mass, labelled a lens, passes near to a source on the plane of the sky.  In the most simple physical description of a stellar microlensing event, the lensed star is considered to be a point source.  In reality, however, the projected stellar disk will be differentially magnified.  This effect has been referred to in the literature as differential limb magnification (DLM).  For DLM to be significant, the source star and lens must be extremely close on the plane of the sky, such that the lens is located on the stellar disk itself or within an angular distance comparable to the star's angular diameter.  Such events have proven useful for observational measurement of limb darkening \citep[see][]{albrow99,albrow01,fields03}.

Not only does the brightness of a stellar disk vary radially, but its spectrum does too.  Current technological limitations dictate that points on a stellar disk cannot be resolved (unless one were observing the Sun).  As such, the integrated spectrum across a star's projected surface is what is measured.  Consequently, the observed spectrum of a star subject to DLM differs from its intrinsic flux spectrum, where absorption line equivalent widths change.  

Rotation of the source has previously been omitted from the discussion of finite-source effects of microlensing in the literature.  To determine the importance of this, we have performed numerical microlensing simulations with DLM, where stellar rotation has been accounted for.  We have analysed these simulations to determine the effect rotation has on deduced stellar properties from observational techniques.  In this paper, we focus on changes in absorption line equivalent widths and briefly assess effective temperature.  As a means of extending the work of \citet{johnson10}, we targeted dwarf stars for these simulations.

For standard stellar spectra, the profiles of absorption lines are symmetric about their peak.  Ordinarily, the effect of stellar rotation is to broaden absorption lines.  If a star is subject to DLM, and assuming the lens does not lie exactly on the star's projected rotation axis, the asymmetrically imposed magnification map on the differentially Doppler-shifted surface causes line symmetry to be broken (see Section \ref{sec:profiles}).

While DLM alters equivalent widths, the added effect of rotation, in reality, does not cause a further alteration; the line's flux is, rather, redistributed.  However, equivalent widths from real data cannot (currently) be measured by direct integration due to finite resolution.  Instead, the standard practice is to fit Gaussian profiles to \sloppy{lines \citep[see, for example,][]{marscher,fulbright06,johnson10,bensby09,bensby10}}.  A Gaussian ceases to be an accurate approximation for a line profile in the above described regime.  With finite-resolution observing tools and noise, the change in line profiles may not be obvious, leaving the \sloppy{equivalent-width-measuring technique unchanged}.  Thus rotation can indirectly affect the \emph{measured} values of equivalent widths.

Initially, this paper was motivated by \citet{cohenpuzzle}, who studied a sample of 16 highly magnified, microlensed dwarf stars in the galactic bulge with high-dispersion spectra.  They described an observed trend between the maximum magnification and measured metallicity of their sample.  The first of several potential explanations for this trend posed by the authors was a finite-source effect, {\frenchspacing{i.e. that}} of DLM.  However, analysis of synthetic spectra by \citet{johnson10} suggested the effect was insufficient to explain the trend.  Recent developments by \citet{bensby13}, though, show this trend to be far less prominent with a larger sample of 58 microlensed bulge stars.

This paper is laid out as follows.  Section \ref{sec:simdesign} covers the design of our simulations and the process of measuring equivalent widths.  Section \ref{sec:runsims} discusses the parameters we chose for the suite of simulations for which we provide results.  Section \ref{sec:results} describes the change to absorption line profiles in our simulation suite, outlines our results from the equivalent width measurements, and assesses the effect of DLM on effective temperature from these simulations.  Finally, we offer concluding remarks in Section \ref{sec:conclusion}.

\section{Simulation Design}
\label{sec:simdesign}

\subsection{Microlensing}

Our simulations were written in \textsc{matlab}.  In brief, the software takes a model spectrum and continuum of a star, coupled with details for some chosen microlensing event, and produces the spectrum that we would theoretically observe, along with the ideal fitted continuum for that star, under the prescribed lens-source geometry.  A full description for an arbitrary microlensing event requires many potential parameters for a simulation.  Angular distances were internally measured in terms of Einstein Ring Radii, $\theta_E$, as this quantity determines the magnification gradient.  Physically, $\theta_E$ is determined by the lens mass and the distances between each of the lens, source, and observer.  The parameters able to be varied in our simulations include: 

\begin{itemize}
\item Radius of the star in Einstein Ring Radii ($R_*$)
\item Separation between the centre of the star and lens in Einstein Ring Radii ($u_*$)
\item Rotation period of the star ($P$)
\item Effective temperature of the star ($T_{\mathrm{eff}}$)
\item Logarithm of the star's surface gravity in cgs units ($\log{g}$)
\item Number of radial points to create interpolated intensity spectra on the stellar surface
\item Angle between the line connecting the centre of the star to the lens in the plane of the sky and the star's rotation axis ($\phi_{\mathrm{lens}}$)
\item Resolving power of the theoretical spectrograph ($\lambda/\Delta\lambda$)
\item Pixel resolution of the theoretical charge-couple device (CCD) detector.
\end{itemize}

The software is designed to use synthetic spectra from the program {\textsc{synthe}}, created by \citet{kurucz}.  We used the Linux version of {\textsc{synthe}} by \citet{sbordone}.  {\textsc{synthe}} produces intensity spectra at a number of equispaced $\mu$-values with a prescribed $T_{\mathrm{eff}}$ and $\log{g}$.  Values of $\mu$ are related to the projected stellar radius fraction, $\rho$, by

		\begin{equation}
		\label{eq:mu}
		\rho = \sin\left(\cos^{-1}(\mu)\right)\ .
		\end{equation}

\noindent  For our simulations, we computed spectra with a sampling resolution of $\Delta\lambda = \lambda/10^6$ with 17 different $\mu$-values.
  
  In order to compute an ``observed" stellar spectrum, numerical integration was performed on the point intensity spectra across the disk.  For this to be accurate, the number of projected radius points was increased using spline interpolation, with a more equispaced grid in $\rho$-space.  We found 200 $\rho$-points was optimal, as with such a grid, fractional movements of the lens toward a grid point did not cause a divergence in the output spectrum.  360 equispaced angular $\phi$-coordinates were also used, providing a circularly symmetric grid of 72 000 points.
  
  Based on the inputs, a Doppler shift at each grid point was first calculated.  For simplicity, the star's rotation axis was always taken to be parallel to the plane of the sky.  Designed to work with dwarf stars, the physical radius of the star was approximated to be proportional to its $T_{\mathrm{eff}}$, using the Sun as a reference.  A magnification for each point, $A_p$, was then calculated using the standard point-source microlensing equation,
  
  	\begin{equation}
		\label{eq:dlm}
		A_p = \frac{u_p^2 + 2}{u_p\sqrt{u_p^2 + 4}}\ ,
		\end{equation}

\noindent	with $u_p$ the angular distance between the point and lens in Einstein Ring Radii.  Each segment of the grid was represented by the average spectrum of its four vertices, weighted by its area.  

We approximated stars as spheres and solid-body rotators in the simulations.  Including differential rotation would, in theory, only produce second-order effects.

In the application of Equation \ref{eq:dlm}, should the lens lie directly on a grid point, an infinite magnification occurs.  In reality, as magnification tends to infinity, the magnified area concerned tends to zero; thus no object, or part thereof, can be infinitely magnified due to microlensing.  To deal with the issue in the code, at the start of each simulation, a check would be run to see if the lens resided exactly on a grid point.  In such an event, the lens would be shifted fractionally away from the point.

\subsection{Equivalent Width Measurement}

An output spectrum from a microlensing simulation represents what would be incident on a theoretical spectrograph.  To simulate observing this with a slit, each output spectrum was convolved with a top-hat function of unit area with a width based on the prescribed resolving power.  Furthermore, the physical detection of such a spectrum would be achieved through a CCD with a finite number of pixels per unit wavelength interval, defining the sampling resolution.  To simulate this binning effect, we applied a further top-hat convolution, again of unit area but of width equal to the pixel width.  The resultant spectrum was subsequently sampled at wavelengths separated by this width to obtain the theoretically observed spectrum.

\begin{figure}[h]
		\centering
		\includegraphics[width=0.48\textwidth]{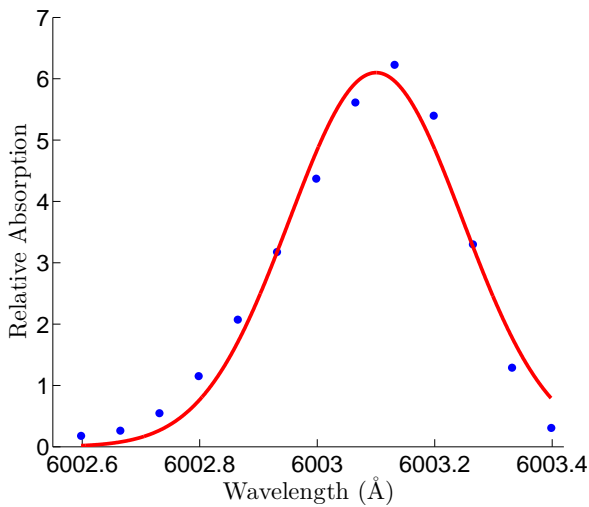}
		\caption{Example of an absorption line from the spectrum of a rotating star subject to DLM from our simulations.  Blue dots are the ``observed'' noiseless data with the red line the fitted Gaussian that is used to determine the equivalent width.}
		\label{fig:gaussfit}
		\end{figure}

In our analysis, to measure the equivalent width of a line, the wavelength range over which the line resides needed to be manually specified.  This section of the spectrum was then subtracted from the continuum and extracted.  A Gaussian profile was subsequently fitted to the extracted data points via a standard non-linear least squares fit, whose area divided by the average continuum flux over the line provided the equivalent width.  Figure \ref{fig:gaussfit} illustrates an example of our fitting procedure.  Note that we do not simulate noise in our analysis.

\section{Running the Simulations}
\label{sec:runsims}

A choice of $T_{\mathrm{eff}}$ and $\log{g}$ should not affect the effect of DLM with rotation, but it may make observing the effect more obvious.  For the purposes of clear absorption lines and desiring a dwarf, we present results only for a model atmosphere of 6500 K with $\log{g} = 4.0$.  The effect should also be more prominent when the star's rotation axis is perpendicular to the lens-source projected separation, allowing for the regions of highest Doppler shift to be preferentially magnified.  As such, we chose to maintain $\phi_{\mathrm{lens}}$ at $\pi/2$ radians.

Each absorption line varies in intensity from the centre to the limb of the star.  Nearer the limb, species responsible for absorption lines lie at a lower average optical depth and, therefore, a lower average temperature.  This hinders absorption by altering the relative populations of excited states.  However, there is also a greater number of absorbers along the line of sight, which increases absorption.  The excitation potential of each line determines which of these competing effects is stronger.  As such, DLM will affect equivalent widths differently for each absorption line, requiring lines with a range of excitation potentials to be assessed to obtain a full picture.

We chose to assess a series of iron lines in the simulated spectra, due to their plentifulness and range of strengths and excitation potentials.  Our linelist, presented in Table \ref{tab:lines}, was formed using the catalogue of \citet{moore} as a guide.  For the purposes of this paper, we categorise lines into ``weak'' and ``mid-strength'', where any line with an equivalent width $<10$ m\AA\ at any point on the 6500-K, $\log{g} = 4.0$ stellar surface is considered ``weak'' and any line with equivalent width $<100$ m\AA\ and no saturation wings is considered ``mid-strength''.  We also provide the intrinsic equivalent-width-to-wavelength ratios of each line in Table \ref{tab:lines} as a quantitative description of strength.

		\begin{table}[h]
		\centering
		\begin{tabular}{c r c c c r}\hline
		$\lambda$ & $W/\lambda$ & Strength & Ion & $\chi$ & $Q$\\
		(\AA) & $\times 10^{-6}$ & & & (eV) & \\\hline
		6003.2 & 12.71 & Mid & Fe I & 3.88 & 1\\
		6024.1 & 17.27 & Mid & Fe I & 4.54 & 0\\
		6185.4 & 0.12 & Weak & Fe II & 3.19 & 2\\
		6187.5 & 0.27 & Weak & Fe I & 2.83 & 2\\
		6240.7 & 3.70 & Mid & Fe I & 2.22 & 2\\
		6436.4 & 0.70 & Weak & Fe I & 4.18 & 0\\
		6446.4 & 1.45 & Weak & Fe II & 6.22 & 1\\
		6810.5 & 4.86 & Mid & Fe I & 4.60 & 0\\
		6837.0 & 1.31 & Weak & Fe I & 4.59 & 0\\\hline
		\end{tabular}
		\caption{Details of the absorption lines measured from the simulations, including: wavelength ($\lambda$); equivalent-width-to-wavelength fraction ($W/\lambda$); intrinsic strength for the model star (see text); ionic species of origin; excitation potential ($\chi$); and $Q$, a quality index describing blending from neighbouring lines, where 0 = no blending, 1 = minimal blending, 2 = noticeable blending.}
		\label{tab:lines}
		\end{table}

A series of simulations were run varying the parameters $u_*$, $R_*$ and $P$.  Typical values of $R_*$ for real microlensing events with dwarf sources are of the order $10^{-3} \theta_E$ \citep[see, for example,][]{choi,shin}.  With an interest in DLM, only values of $u_*$ close to that of $R_*$ were used.  Two values of $R_*$ were used for the grid of simulations.  For each of these, 32 values of $u_*$ with a fractional dependence on $R_*$ were simulated, such that the events covered the lens at positions on the disk of the star and just off the disk within a few angular stellar radii.  An emphasis was placed on events close to the limb, where the greatest effect on spectra occurs.  A non-rotating case and two rotating cases were taken for each of these also: one slow rotator with a period of 30 days and a fast rotator of period 5 days, giving 192 simulations in total.  A period of 5 days is unusually fast for an aged (non-binary) dwarf star, but can occur (as can shorter periods) in young clusters \citep[see, for example,][]{hartman10,agueros11,irwin11,meibom09,meibom11} and provides a case that should emphasise any present effect.  The 30-day-period case was chosen as a value slightly slower than that of the Sun, based on \citet{snodgrass}.  The non-rotating case is equivalent to observing a star whose rotation axis is perpendicular to the plane of the sky; therefore, it provides a physically viable situation and a control to show the added effect of rotation to DLM.

The true equivalent widths of the lines were first measured from the intrinsic flux spectrum using the same Gaussian-fitting method as the simulation output spectra for a reference.  A resolving power of 30 000 was chosen for the theoretical spectrograph with a sampling resolution of $\Delta\lambda = \lambda/90\ 000$.

\section{Results}
\label{sec:results}

\subsection{Line Profiles}
\label{sec:profiles}

\begin{figure}[h!]
\centering
\includegraphics[width=0.48\textwidth]{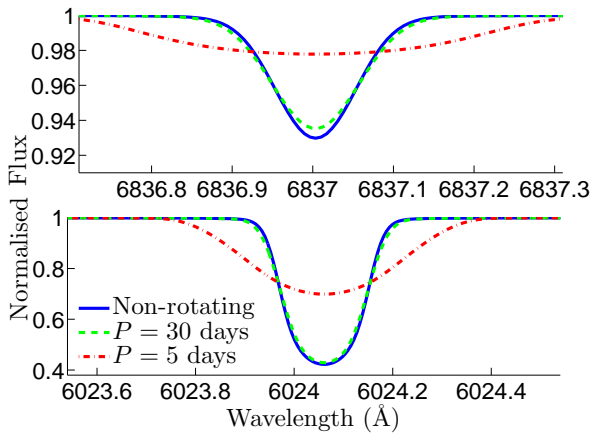}
\caption{Non-lensed, integrated line profiles for various rotation periods of the same synthetic stellar spectrum as processed through the simulations.}
\label{fig:nonlensprofs}
\end{figure}

As a test of the simulations and a reference for the strength of rotational broadening, Figure \ref{fig:nonlensprofs} provides line profiles for the 6024-\AA\ and 6837-\AA\ lines for each of the non-, slow-, and fast-rotating cases, where no lensing has occurred.

\begin{figure}[h!]
		\centering
		\includegraphics[width=0.48\textwidth]{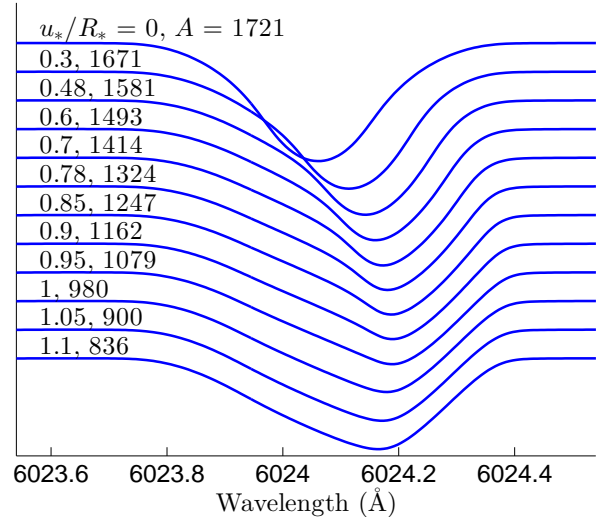}
		\caption{Output line profiles (manually separated) from the relative-flux spectra of the simulations with \textit{R}$_* =$ 0.00125$\theta_E$ and \textit{P} $=$ 5 days, labelled with the lens position, $u_*/R_*$, and total magnification factor, $A$.  Distortion to the profile increases as regions of higher Doppler shift are preferentially magnified.}
		\label{fig:profile}
		\end{figure}

As discussed in Section \ref{sec:intro}, and evident from the discrepancy between the sampled and fitted profiles in Figure \ref{fig:gaussfit}, the effect of rotation with DLM is to break absorption line symmetry.  Figure \ref{fig:profile} displays how the profile of a typical line changes for microlensing events of varying $u_*$.  The peak of the line is shifted by the greatest amount when the lens is at the limb, as the point of greatest redshift (or blueshift, had the geometry been reversed) is preferentially magnified.  As the lens is shifted further off the limb, the magnification gradient across the disk becomes less steep, so the peak of the line begins to tend back to its intrinsic position.  The pronounced distortion in the line profiles highlights how an equivalent width measurement based off a Gaussian fit generates error.

\subsection{Equivalent Widths}
\label{sec:EWs}

\begin{figure*}[ht!]
		\centering	
		\includegraphics[width=\textwidth]{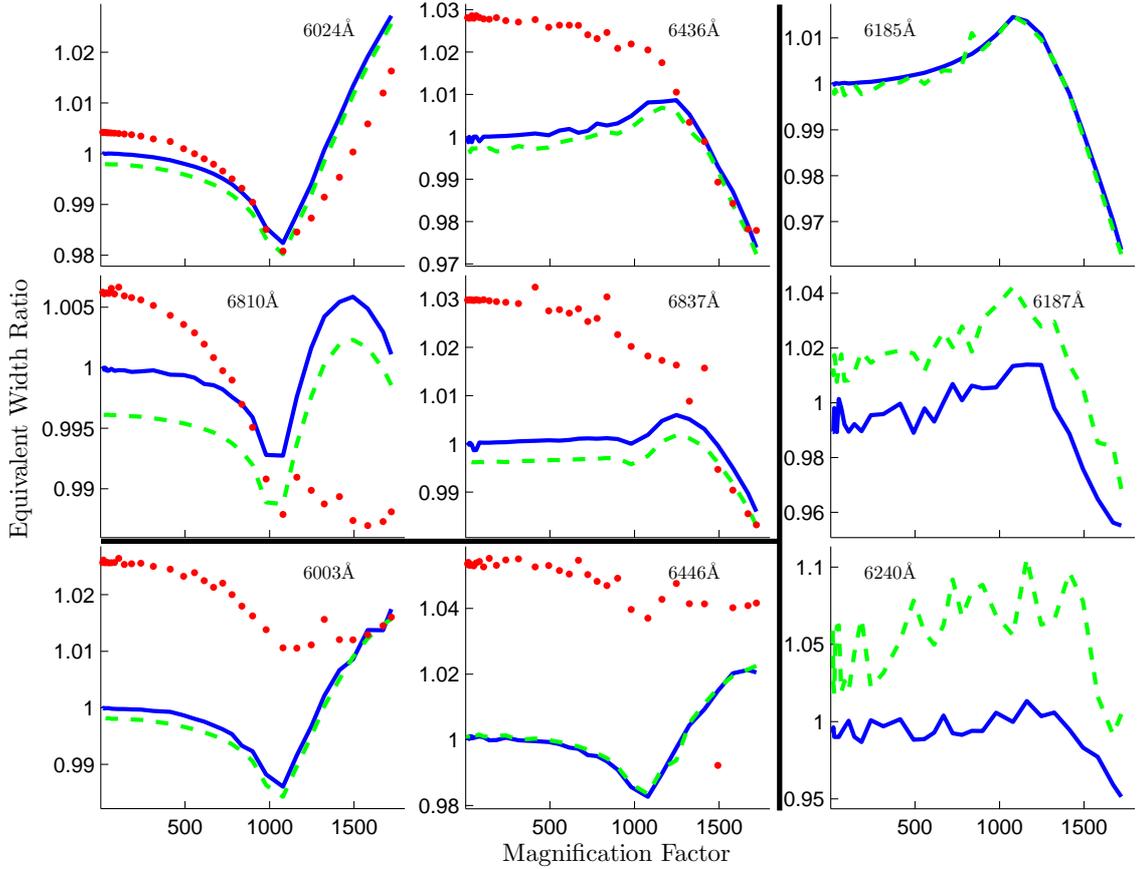}
		\caption{Equivalent width measurements given against the total magnification of the star from each simulation where $R_* = 0.00125\theta_E$.  The blue, solid lines indicate the simulations without rotation, providing the control for the raw effect of DLM. The green, dotted lines and red dots correspond to the simulations with a stellar rotation period of 30 days and 5 days, respectively.  A magnification of 980 corresponds to the lens lying on the limb.  The 4 plots in the top-left correspond to the $Q=0$ lines, while the 2 in the bottom-left have $Q=1$, and the 3 on the right have $Q=2$.  5-day-rotation data for the $Q=2$ lines were unreliable, as blending became an issue.  Plots in the left column are mid-strength lines, while the remainder (except the 6240-\AA\ line) are weak.}
		\label{fig:EWs}
		\end{figure*}

We present the results in plots of magnification factor against the ratio of each equivalent width to its intrinsic value in Figure \ref{fig:EWs}.  We provide the results from the simulations of $R_* = 0.00125\theta_E$ only, as the results from the $R_* = 0.005\theta_E$ simulations were almost identical but for a scaled magnification axis.

Figure \ref{fig:EWs} shows that stellar rotation indeed has an effect on the \emph{measured} values of equivalent widths by Gaussian fit.  In the slow-rotation regime, the contribution of rotation to the raw DLM effects is consistently minimal.  Only for the fast rotators does the effect become noticeable, and, in some cases, dominant.  Even with a short rotation period for dwarfs of 5 days, the measured equivalent widths are only altered by $<6\%$.  For the $Q=2$ lines in the fast-rotation regime, blending overtook DLM as the primary reason for a change in measured equivalent width.  Consequently, Figure \ref{fig:EWs} omits the fast-rotation data for these lines.  This was not an issue in the other cases, being only responsible for the small scatter seen in the equivalent width measurement.

 The highest magnification events correspond to the lens being positioned at the source's centre.  When the lens lies on the source's disk, the equivalent widths of all lines in the observed spectrum will be very close to the values of the intensity spectrum at the point of preferential magnification.  The competing centre-to-limb variations, described in Section \ref{sec:runsims}, give rise to the structured curves seen in Figure \ref{fig:EWs}.
 
 The results for the most isolated lines ($Q = 0$) consistently show the slow-rotating events to provide smaller measurements of equivalent width at all magnifications. Meanwhile, the fast-rotating cases give higher equivalent widths at lower magnifications, with similar but lower equivalent widths when strongly magnified (when the lens lies on the source's disk).  For the weak lines, the measured equivalent widths tend to be $3\%$ above their intrinsic value as the magnification factor approaches 1, while the mid-strength lines indicate a difference of $\sim0.5\%$.  This is purely a rotational effect, where the weaker lines' height-to-width ratio is increased by a greater amount due to rotational broadening.  The fitting of a non-Doppler-broadened Gaussian to the profile hence induces a greater change in measured equivalent width for the weaker lines.  The remaining difference between the non- and fast-rotating cases are caused by the asymmetric nature of the line profiles.

\subsection{Effective Temperature}

\begin{figure}[h!]
		\centering
		\includegraphics[width=0.48\textwidth]{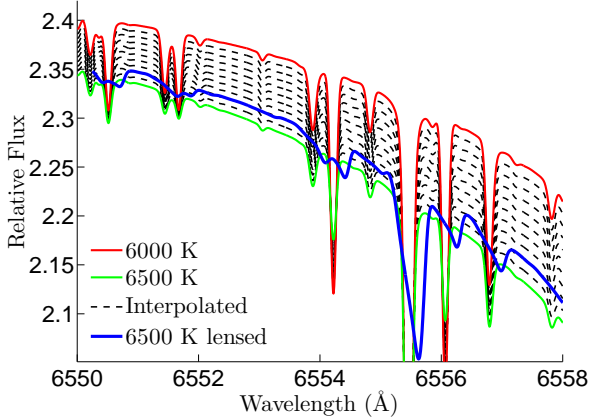}
		\caption{The red wing of H$\alpha$ for various model spectra.  Top-most, red, solid line: Intrinsic flux spectrum from a 6000-K, $\log{g} = 4.0$ star. Bottom-most, green, solid line: As for red but with $T_{\mathrm{eff}} = 6500$ K.  Black, dashed lines: Interpolated spectra between red and green in steps of 50 K.  Thick, blue, solid line: Output microlensed spectrum for a 6500-K, $\log{g} = 4.0$ star with simulation parameters $R_* = u_* = 0.00125\theta_E$, $P = 5$ days.  The simulation output spectrum aligns very well with the H$\alpha$ wing for the $T_{\mathrm{eff}} = 6400$ K model and displays the distortion effect of DLM with rotation on the weaker, superimposed absorption lines.}
		\label{fig:Halphawing}
		\end{figure}

Should one use equivalent width measurements as in Section \ref{sec:EWs} to derive an elemental abundance, a negligible error would be carried through.  However, a measurement of $T_{\mathrm{eff}}$ would also be required.  One simple method of measuring the $T_{\mathrm{eff}}$ of a star is to compare the wings of Balmer lines to those of model spectra. \citet{johnson10} showed that DLM could cause a difference between the measured and true $T_{\mathrm{eff}}$ by approximately 100 K via this method.  Here, this method is repeated with the inclusion of rotation by comparing the wings of H$\alpha$. The most extreme microlensing event has been considered ($R_* = u_* = 0.00125\theta_E$, $P = 5$ days) and the output spectrum of this simulation compared with intrinsic flux spectra from the \citet{kurucz} models.  The extreme broadness of H$\alpha$ means the distortion to its shape due to the DLM-rotation combination will be minimal compared with weak or mid-strength lines.  Figure \ref{fig:Halphawing} displays the output spectrum, centred on the red wing of H$\alpha$, of this event, with intrinsic flux spectra for the same star (6500 K) and a 6000-K star (also with $\log{g} = 4.0$) plotted along with linearly interpolated spectra between them in steps of 50 K.
		
The broadening of H$\alpha$ caused by rotation is on a far smaller scale than the width of the line.  As a result, there is negligible difference between the wings of H$\alpha$ for the fast- and non-rotating lensed spectra.  The simulation output spectrum falls almost perfectly on the interpolated 6400-K model at the H$\alpha$ wings in Figure \ref{fig:Halphawing}.  Thus we have reaffirmed the result that DLM can alter a measured $T_{\mathrm{eff}}$ by 100 K, and we conclude that rotation has no additional effect on this difference.

The lensed spectra in Figure \ref{fig:Halphawing} have not been convolved; they are the H$\alpha$ profiles that would be incident on a spectrograph slit.

\section{Conclusion}
\label{sec:conclusion}

During instances of stellar microlensing when the lensing object lies extremely close to or on the disk of the star in the plane of the sky, finite-source effects cease to play a negligible role in the interpretation of stellar spectra.  DLM, the effect of a magnification gradient across a projected stellar surface, alters observed spectra in two ways. First, flux spectra appear more like the intensity spectra from the area of preferential magnification.  Second, the combination of this with a differentially Doppler-shifted surface induces distortion in absorption line profiles.  For real data, line profiles typically are not resolved, requiring an analytic fit of the profile to measure equivalent widths.  These are assumed to be symmetric and approximately Gaussian in nature, which is no longer true for a rotating star subject to DLM.  To determine the extent of these effects on equivalent width and $T_{\mathrm{eff}}$ measurement, we have presented results from microlensing simulations using a modelled dwarf star atmosphere.

From the simulation outputs of a 6500-K, $\log{g} = 4.0$ star, we measured the equivalent widths of nine iron lines from a series of microlensing geometries.  Predictably, for the simulations of non-rotating stars, a maximum or minimum measured equivalent width corresponds to a microlensing event where the lens was very close to the limb, dependent on how the line's strength varied with projected radius.  The addition of a slow rotation period caused a displacement in the measured equivalent widths, but still maintained the same trend with magnification.  For a ``slow'' rotator like the Sun, the distortion to spectral lines would be minimal, so a Gaussian fit would remain reasonable.  The inclusion of a short rotation period instead produced a different trend and, in most cases, more extreme values of measured equivalent width.  Although we have shown rotation to induce an effect on measured equivalent widths, we reason its contribution could only account for variations in measurements of, at most, a few percent with decent spectra.

Finally, based on comparisons of profiles of H$\alpha$, it was confirmed that DLM can generate $T_{\mathrm{eff}}$ measurements 100 K cooler than a star's true value, using the same model atmosphere with microlensing parameters $R_* = u_* = 0.00125\theta_E$.  This value is in accordance with similar results found by \citet{johnson10}.  Including a rotation period of 5 days provided an identical $T_{\mathrm{eff}}$ measurement.  It is hence concluded that stellar rotation should not affect the derived $T_{\mathrm{eff}}$ of a star from the H$\alpha$ profile method.

\section*{Acknowledgements}

ARHS would like to thank Professors Peter Cottrell and John Hearnshaw at the University of Canterbury for discussions on the topics of stellar atmospheres, line profiles and abundance analysis, that provided insight for this project.

%\end{multicols}

\end{document}